\documentclass[aps,prl,graphicx,twocolumn]{revtex4-1}
\draft % marks overfull lines with a black rule on the right
\usepackage{graphicx}
\usepackage{epstopdf}
\usepackage{natbib}

\begin{document}

%\preprint{}

\title{Rhodamine B as a probe for phase transition in liquid solutions} %Title of paper



\author{I. Shishkin}
\email{ivanshishkin@post.tau.ac.il}
%\thanks{}
\affiliation{School of Electrical Engineering, Tel Aviv University, Ramat-Aviv, 69978, Israel}
\author{T. Alon}
\affiliation{School of Electrical Engineering, Tel Aviv University, Ramat-Aviv, 69978, Israel}
\author{R. Dagan}
\affiliation{School of Electrical Engineering, Tel Aviv University, Ramat-Aviv, 69978, Israel}
\author{P. Ginzburg}
\affiliation{School of Electrical Engineering, Tel Aviv University, Ramat-Aviv, 69978, Israel}
\affiliation{ITMO University, St. Petersburg 197101, Russia}


%\date{\today}

\begin{abstract}
Local environment of fluorescent dyes could strongly affect emission dynamics of the latter. In particular, both signal intensities and emission lifetimes are highly sensitive to solvent temperatures. Here, temperature-dependent behavior Rhodamine B fluorescence in water and ethanol solutions was experimentally investigated. Phase transition point between liquid water and ice was shown to have a dramatic impact on both in intensity (30-fold drop) and in lifetime (from 2.68 ns down to 0.13 ns) of the dye luminescence along with the shift of spectral maxima from 590 to 625 nm. At the same time, ethanol solvent does not lead any similar behavior. The reported results and approaches enable further investigations of dye-solvent interactions and studies of physical properties at phase transition points.
\end{abstract}


\maketitle %\maketitle must follow title, authors, abstract and \pacs

\section{Introduction}
Temperature is one of the most important macroscopic measures, characterizing properties of surrounding environment. The ability to control and monitor the temperature at real time and with high precision is required in vast number of applications. Several techniques of micro- and nano-thermometry \cite{Zhou2016} have been developed in order to monitor properties of very small objects. In particular, inapplicability of conventional methods, such as thermocouple measurements or infrared thermography for temperature studies on micro- and nano- scales, has led to search for alternative probes. Recently, the methods basing on the change of fluorescent properties of the quantum emitters have been implemented. Generally, luminescent techniques can be divided in two broad categories: the methods, which are based on relative change of emission intensity of quantum tags \cite{Zohar1998,Ebert2007,Schaerli2009,Ross2001} and the ones, relying on the change of the lifetimes \cite{Benninger2005, Muller2009, Bennet2011}. The lifetime-based techniques generally require more complex diagnostic equipment compared to intensity-based measurements, though they are less sensitive to environmental effects. Besides the temperature measurements, the fluorescent techniques also were employed in studies of flows inside micro channels in lab-on-chip applications \cite{Benninger2006, Haro-Gonzalez2012}, in microviscosity investigations \cite{Yamaguchi2012, Someya2010}, airflow sensing\cite{Wohl2015,Gallery1994}, and as pressure measurements \cite{Johann2015}. In order to achieve stable and reliable results, the certain balance between isolation and overlap of the probe with the environment is desired. This could be achieved, for example, by embedding dyes in polymer or inorganic host matrix to control their interaction with surrounding environment \cite{Chauhan2014, Duong2007}.
Phase transitions sensing in a host media is one among other interesting applications of molecular probes. It has been demonstrated earlier, that phase transitions can be detected by photoluminescence of embedded ions in crystalline solids \cite{Townsend2002, Townsend2008} and with dyes in liquid crystals \cite{Subramanian1982a}. Recently it has been demonstrated, that number of quantum emitters exhibit sensitivity to phase transitions occurring in solvents. In the case of CdSe/ZnS quantum dots \cite{Antipov2011} it has been reported, that photoluminescence intensity demonstrates significant growth close to a phase transition point, and accompanied by shift and linewidth increase of the emission peak. Quantum dots also demonstrate a drop in lifetime at the transition point. The studies on freezing of fluorescent proteins (phycobiliproteins) in aqueous media also demonstrate  sensitivity to solvent phase transition\cite{Maksimov2012}. It was shown, that the fluorescence lifetime of the proteins strongly depends on the freezing rate. In the case of slow freezing the protein solution exhibits the drop of the lifetime from 1.8 ns to 200 ps, which is explained by photoisomerization of proteins into a quenching form. In the case of rapid freezing no significant change in the lifetimes was observed, underlining the importance of the transient processes.
 
Rhodamine B (RhB) is a well-known laser dye with high quantum yield\cite{Drexhage1976}, which is commonly used as fluorescence lifetime standard. Rhodamine dyes are frequently used as microprobes due to their relatively high sensitivity to the temperature change\cite{Ross2001, Samy2008}. Nevertheless, their use as probes has certain drawbacks, first, the fluorescence lifetime depends on the concentration of the dye – fluorescence quenching takes place at high concentrations \cite{Arbeloa1989} along with presence of dual-exponential decay due to formation of non-fluorescent aggregates\cite{Arbeloa1988,Kemnitz1991}. Second, the dyes can exhibit photobleaching due to interaction with dissolved oxygen or at high excitation powers. All those processes could be affected by changes in host environment. 
In this work the behavior of luminescence of Rhodamine B dye at the phase transition temperatures of two commonly used solvents – ethanol and water was analyzed. In particular, the existence of a significant drop in lifetime and photoluminescence (PL) intensity in case of water solutions was observed. In the case of ethanol solutions, temperature-induced changes in fluorescence lifetimes did not show critical discontinuities at the phase transition point.
\section{Results}
The decay dynamics and luminescence intensity of fluorescent dyes strongly depends on properties of ambient environment of the molecules, such as environment temperature, polarity, viscosity, pressure and phase transitions in the medium \cite{Lakowicz_book}. The thermal sensitivity could be described by the following relation: 
\begin{equation}
S = \frac{\Delta Q}{(Q_{T} \Delta T)}\times 100\%
\end{equation}
where S is the sensitivity measure $[\%^{\circ}C^{-1}]$, Q$_{T}$ is a reading (measurable) at initial temperature, $\Delta Q$ corresponds to the change in the signal over the temperature span $\Delta T$. In the forthcoming experiments $\Delta Q$ corresponds to either lifetime changes or light intensities at the emission peak.
\begin{figure}
	\centering
	\includegraphics[width=1.0\linewidth]{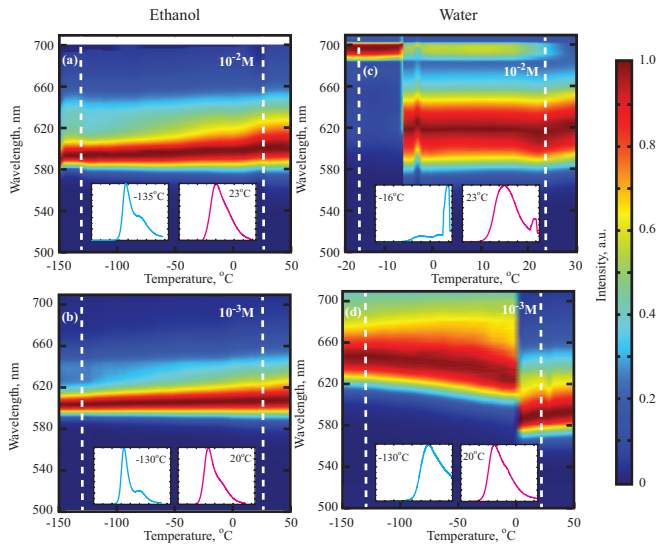}
	\caption{\label{fig:PLMAP} Photoluminescence spectra of Rhodamine B solutions at different concentrations (rows) and solvents (columns). (a) ethanol, $10^{-2}$M concentration. (b) ethanol, $10^{-3}$M concentration (c) water, $10^{-2}$M, (d) water, $10^{-3}$M. All measured spectra were normalized to their maximum for each temperature point. Insets: slices of the spectra at the places, indicated with white dashed lines on the color maps.}
\end{figure}
The measured photoluminescence spectral data over broad range of temperatures and two different concentrations of dyes appears on \ref{fig:PLMAP}. Each presented spectrum at a given temperature was normalized to its maximum, since the absolute luminescence signal intensities strongly depends on the temperature and exhibit significant drop at the freezing of water, as will be discussed hereafter.

In case of ethanol solutions with $10^{-3}$M and $10^{-2}$M concentrations a single-band luminescence was observed, though the spectra are shifted to longer wavelengths in case of higher concentration. The luminescence peak occurs at 600 nm for $10^{-2}$M RhB ethanol solution and 582 nm for $10^{-3}$M solution at $20^{\circ}$C. With the decrease of the temperature down to $-150^{\circ}$C the solution with higher concentration exhibits shift of spectral maxima down to 594 nm while $10^{-3}$M solution does not exhibit significant spectral shift. Both spectra demonstrate linewidth narrowing with the decrease of temperature. The intensity of the luminescence is also strongly dependent on the temperature. No spectral features were observed at ethanol freezing temperature ($-114^{\circ}$C) for both solutions of dye in ethanol. In the temperature range from 10 to $50^{\circ}$C we found that sensitivity of RhB in ethanol is equal to $1.75\%^{\circ}$C$^{-1}$.
\begin{figure}
	\centering
	\includegraphics[width=1.0\linewidth]{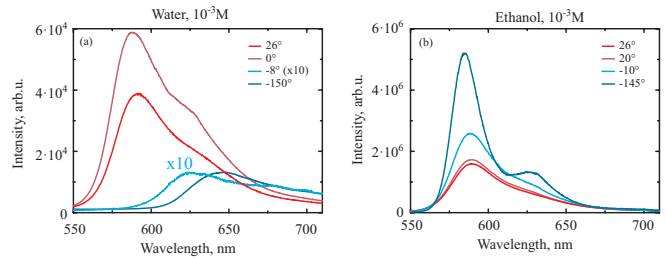}
	\caption{\label{fig:PLspec} Photoluminescence spectra at different temperatures of Rhodamine B solutions in (a) water (b) ethanol.}
\end{figure}

The spectra of aqueous solutions of dye at room temperature have luminescence emission maximum at 590 nm ($10^{-3}$M in water) and 615 nm ($10^{-2}$M) respectively. One can observe an additional spectral peak at 697.5 nm in $10^{-2}$M solution at room temperature ($20^{\circ}$C) which therefore did not appear in case of less concentrated solution. Being concentration dependent, this peak is the signature of collective behavior of the dye molecules in the solution. This effect could be attributed to formation of dye aggregates \cite{Choi2014a,Setiawan2010,Wurthner2011}, nevertheless spectral measurements of this kind only provide an indirect signature. Remarkably, the freezing of dye solution appears in spectra as a rapid drop in the intensity of band at 590 nm ($10^{-3}$M) and 615 nm ($10^{-2}$M) respectively. In case of $10^{-3}$M solution, the luminescence signal peak shifts to 625 nm along with significant drop in the intensity. The 697.5 nm band luminescence intensity appears to remain unchanged when the solution freezing point is passed. More detailed luminescence spectra are presented on Fig.\ref{fig:PLspec}.
The intensity-based sensitivity of aqueous Rhodamine B solution yields values of $1.6\%^{\circ}$C$^{-1}$ for the temperatures from 10 to $50^{\circ}$C.

\begin{figure}
	\centering
	\includegraphics[width=1.0\linewidth]{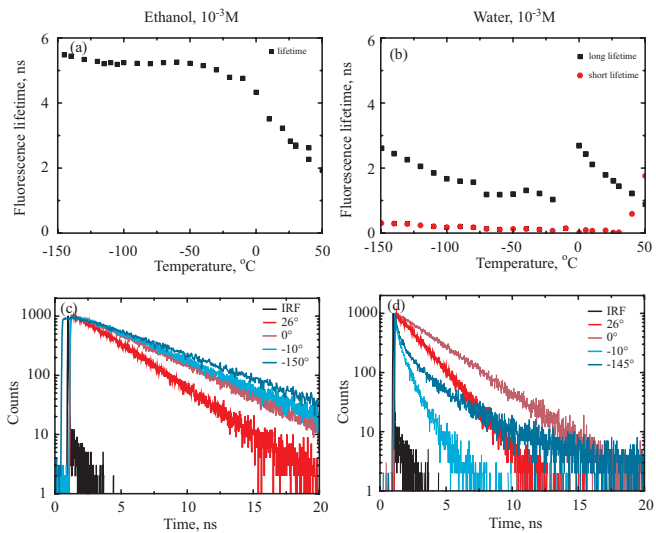}
	\caption{\label{fig:Lifetimes} Measured lifetimes of Rhodamine B dyes, dissolved in ethanol and water solutions. (a, b) – lifetimes over temperature plots. (a) – single exponential fit; (b) double-exponential fit. (c-d) time-dependent decay curves for different temperatures in $^{\circ}$C (in legends).}
\end{figure}

To further investigate the changes occurring at phase transition point of water, we have measured the lifetimes of solutions at different temperatures. The results are presented on Fig.\ref{fig:Lifetimes}. Lifetime data was recorded with monochromator position set to the maximum intensity of measured PL spectra at each temperature point.

The measured time-dependent data and extracted lifetimes for Rhodamine B in ethanol are presented in Figure \ref{fig:Lifetimes}(a,c). The data appeared to be well-described by single-exponential decay. The measured lifetime of $10^{-3}$M dye solution in ethanol at $20^{\circ}$C was found to be $\tau=3.23$ ns. This value differs  from the data reported in \cite{Kristoffersen2014} ($\tau=2.68$ ns) which could be attributed either to inaccuracies in measurements of temperature or by local changes of concentration. The lifetime demonstrates growth down to T = $-50^{\circ}$C where the lifetime was found to be $\tau=5.22$ ns. Upon further decrease of the temperature no significant changes in the lifetime were observed and the lifetimes were found to be $\tau=5.48$ ns at $-145^{\circ}$C. In the temperature range from 50 to $-50^{\circ}$C the dye solution in ethanol provides slope of $(d\tau/dT)=-(0.041\pm 0.002)ns/^{\circ}$C, which corresponds to sensitivity of $0.78\%^{\circ}$C$^{-1}$.

The solution of Rhodamine B in water with concentration of $10^{-3}$M can be described by single-exponential decay at temperatures above freezing point (see respective curves on Fig. \ref{fig:PLspec}(d)). The lifetime at T = $20^{\circ}$C is found to be $\tau=1.79$ ns, which is in agreement, than reported earlier \cite{Boens2007}. The lifetimes of aqueous solution of the dye demonstrates an increase in the fluorescence lifetime with the cooling of solution down to the freezing point up to the value of $\tau=2.68$ ns at T = $0^{\circ}$C. The transition through the freezing point can be distinguished by a rapid drop in luminescence intensity along with the drop in the lifetimes down to $\tau_{1}=0.13$ ns and $\tau_{2}=0.12$ ns respectively (measured at T = $-10^{\circ}$C). With further decrease of temperature of the sample one can observe two different behaviors of lifetimes: one of them remains relatively short (on order of $\tau_{1}=0.31$ ns at T = $-150^{\circ}$C), while the other one demonstrates an increase (up to $\tau_{2}=2.61$ ns at $-150^{\circ}$C). In range of temperatures from 0 to $50^{\circ}$C the aqueous solution provides the sensitivity of $(d\tau/dT)=-(0.035\pm 0.002)ns/^{\circ}C$, resuling in the values of sensitivity of $1.34\%^{\circ}$C$^{-1}$. In the phase transition point the lifetime exhibits drop from 2.68 ns down to 0.13 ns, which could be considered as 2.55 ns/$^{\circ}$C sensitivity at transition point.

\begin{table}
	\caption{\label{tab:table1}Comparison of relative temperature sensitivity of different methods in different solutions}
	\begin{ruledtabular}
		\begin{tabular}{ccc}
		& \multicolumn{2}{c}{Sensing method} \\ 
		Solvent & Lifetime-based & Intensity-based\\
			\hline
			Ethanol &$0.78\%^{\circ}$C$^{-1}$ & $1.75\%^{\circ}$C$^{-1}$\\
			Water & $1.34\%^{\circ}$C$^{-1}$ & $1.6\%^{\circ}$C$^{-1}$ \\
			
		\end{tabular}
	\end{ruledtabular}
\end{table}

\section{Outlook and conclusions}
Rhodamine B as phase transition sensor (e.g. freezing of water) has been demonstrated. It was shown, that fluorescent properties are both very sensitive to the temperature of the host solution and has a dramatic dependence on the solvent type. The major differences between the solvents manifest themselves at the phase transition points, there both emission spectra and lifetimes have abrupt changes. Specifically, it was found that the water freezing is accompanied by rapid drop in luminescence intensity along with the shift of luminescence peak from 590 nm to 625 nm along with drop in lifetime from 2.68 ns down to 0.13 ns. At the same time, ethanol dye solutions did not exhibit any discontinuities at its phase transition point. Furthermore, performance of Rhodamine B as a temperature probe was assessed. Two methods for temperature sensing were compared, namely intensity and lifetime. It was shown, that lifetime methods exhibit generally lower sensitivity compared to intensity-based methods away from the phase transition point (1.6\% $^{\circ}$C$^{-1}$ for intensity-based method compared to 1.34\% $^{\circ}$C$^{-1}$ for lifetime-based method in case of aqueous solution of Rhodamine B). The data on the sensitivities is summarized in Table \ref{tab:table1}. Nevertheless, a dramatic drop both in intensity and lifetime in case of aqueous solution could be considered as sensitive probe for phase transition in water. From the fundamental standpoint, the reported results and approaches enable further investigations of dye-solvent interactions and studies of physical properties of liquids at phase transition points. Application-wise, properly chosen fluorescent tags could enable efficient monitoring of dynamical processes in liquid solutions.
\section{Methods}
\textit{Samples preparation.} Solutions of Rhodamine B (fluorescence-grade, 83689 Sigma-Aldrich) were prepared using deionized water (Millipore MilliQ) and absolute ethanol (Bio-Lab, analytical grade). The solutions with concentrations of 10$^{-2}$M and 10$^{-3}$M were studied in this work. The dye solutions were loaded in air-tight cell machined in stainless steel plate and covered with sapphire window with the cell thickness of 50 micrometers. The cell was placed inside the cryostat to prevent the condensation of water vapors during low-temperature measurements.

\textit{Experimental Setup.} The photoluminescence and lifetime measurements were taken using Time Resolved Photo-luminescence (TRPL) technique. The Ti:Sa laser (Tsunami, Spectra Physics) is pumped by CW Nd:YAG laser (Millenia, Spectra Physics) provides short pulses of width about 1 ps with the repetition rate of 80 MHz. The laser repetition rate is reduced using a pulse picker (model 3980, Spectra Physics, Inc.) to the rate of 4 MHz and the laser wavelength is doubled using a nonlinear crystal up to 404 nm. The laser beam power and spot size are adjusted to prevent crystal heating and surface damage. The measurement cell is installed in a high vacuum cryostat chamber (VPF-800, Janis Research Co.), on a sample holder that can be cooled or heated between 77 K and 800 K. The photo-luminescence (PL) emitted from the sample is collected by a combination of a subtractive double monochromator (Triax 190, Jobin Yvon Inc.), an avalanche photo diode (APD) single photon detector (ID-100, IDQ, Inc.) and a Time-Correlated Single Photon Counter (T900, Edinburgh Instruments). The overall instrument response (full width at half maximum) is about 40 ps. The excitation beam is slanted in respect to cell surface in order to minimize the effects of reflection in the cell. A long-pass filter with cutoff wavelength of 500 nm was used to filter the excitation emission. A set of neutral density filters was used to avoid detector saturation during photoluminescence and TCSPC measurements.

\section{Acknowledgements}
The authors thank colleagues from Tel Aviv University - Gary Rozenman and Prof. Tal Schwartz for the help with calibration samples, Yefim Yankelevich for the help with design and fabrication of measurement cell and Prof. Yossi Rosenwaks for providing the experimental setup. This work was supported, in part, by TAU Rector Grant and German-Israeli Foundation (GIF, grant number 2399).


\end{document}